# Quasiparticle scattering in three-dimensional topological insulators near the thickness limit

Haiming Huang[1†], Mu Chen[2†], Dezhi Song[1], Jun Zhang[1*], Ye-ping Jiang[1*]

*1 Key Laboratory of Polar Materials and Devices (MOE) and Department of Electronics, East China Normal University, Shanghai 200241, China*

*2 Beijing WeLion New Energy Technology Co., Ltd, Beijing 102402, China*

In the ultra-thin regime, $Bi_2Te_3$ films feature two surfaces (with each surface being a two-dimensional Dirac-fermion system) with complicated spin textures and a tunneling term between them. We find in this regime that the quasiparticle scattering is completely different compared with the thick-film case and even behaves differently at each thickness. The thickness-dependent warping effect and tunneling term are found to be the two main factors that govern the scattering behaviors. The inter-band back-scattering that signals the existence of a tunneling term is found to disappear at 4 quintuple layers by the step-edge reflection approach. A four-band model is presented that captures the main features of the thickness-dependent scattering behaviors. Our work clarifies that the prohibition of back-scattering guaranteed by symmetry in topological insulators breaks down in the ultra-thin regime.

**Keywords:** three-dimensional topological insulator, surface states, thickness limit, scattering

[†] These people contribute equally to this work.

* Corresponding authors. Email: zhangjun@ee.ecnu.edu.cn, ypjiang@clpm.ecnu.edu.cn

The way electrons scatter off crystal defects governs the transport properties of materials, especially at low temperatures where the coherent propagation of electron leads to strong interference in the inevitable presence of various kinds of scatterers in real materials. The scattering behaviors of electrons are extremely sensitive to fermi surface contours, spin or pseudo-spin textures, especially in two-dimensional (2D) systems [1-4]. The surface states (SSs) of a three-dimensional (3D) topological insulator (TI) form a 2D Dirac fermion system with a helical spin-momentum texture [5-7], preventing backscattering and embeds a nontrivial Berry's phase of $\pi$ [8-10], leading to the weak-antilocalization phenomenon [11-14].

In ultra-thin TI films, the scattering behavior might be even more complicated. The system features two SSs (top and bottom) and a tunneling term between them [15-24]. The two SSs with opposite helicity are related by inversion symmetry and overlap in real space in the presence of inter-surface tunneling, in which case the backscattering is no longer forbidden and new interference patterns might appear. Here we propose a quasi-particle interference (QPI) approach based on scanning tunneling spectroscopy (STS) to investigate the crossover phenomena in QPI near the critical thickness. Nonetheless, the experimental protocol (impurity doping) that is used to introduce point scatterers [3] for bulk samples or thick films fails in the ultra-thin film case. The much-reduced carrier density and the hybridization gap in the SSs lead to the ineffective electron screening compared with the thick-film case. The introduced impurities will cause potential fluctuation that is fatal to QPI. This dilemma can be circumvented by taking the step edge as the scattering source that acts as a 1D reflector and also embeds point scatterers because of lattice imperfections along the edge. No impurities are introduced into the film to keep the uniformity of surface potential.

$Bi_2Te_3$ is a typical strong 3D TI, having a single Dirac conic structure in the SSs and a relatively large bulk gap [5,25]. In addition, among the TI family of Bi(Sb)-Se(Te), the QPI of SSs in $Bi_2Te_3$ is greatly enhanced because of the strong hexagonal warping [26]. Hence, ultra-thin $Bi_2Te_3$ films are the best candidate to investigate possible crossover phenomena in scattering behaviors of SSs near the criticality. We find that the QPI patterns featuring the back-scattering process start to appear and gradually dominate the scattering behaviors with the decreasing thickness. The critical thickness is found to be 4 quintuple layers (QL). The experimental QPI patterns for films of 1-4 QLs agree well with the simulations. Our results provide direct and microscopic origin of the crossover phenomenon in transport behaviors of ultra-thin TIs.

Ultra-thin films of $Bi_2Te_3$ of various thicknesses were prepared by molecular beam epitaxy on the graphene/SiC(0001) substrate, which provides a uniform electrostatic background that doesn't introduce extra potential variations into the films. Figure 1(a) shows the surface topography image of a thick (~ 30 QL) $Bi_2Te_3$ film with Sb adatoms. The QPI pattern induced by these surface impurities is shown in the dI/dV map of Fig. 1(b) taken at 300 meV. The Dirac energy for this film is ~ -250 meV (Fig. S5(f) of [27]). The QPI pattern in the reciprocal space in Fig. 1(c) is obtained by Fast Fourier transformation (FFT) of Fig. 1(b). Figure 1(d) is the calculated QPI as described later. The strong hexagonal warping effect enhances the scattering vectors connecting parallel sectors (along $\bar{\Gamma} - \bar{M}$, as shown in the insert of Fig. 1(d)) on the constant energy contour (CEC) [8]. This leads to scattering waves along $\bar{\Gamma} - \bar{M}$ in the real space QPI image (Fig. 1(b)) as well as scattering arcs in the reciprocal one (Fig. 1(c)).

In contrast, the QPI for ultra-thin films behave very differently compared with that of thick films. The real and reciprocal QPI patterns for the 1-4 QL films at representative energies are shown in Figs. 1 (e)-(l) (see Figs. S1-S4 in [27] for QPI images at other energies). In the 1-QL case, the QPI pattern in the reciprocal space becomes a circle. In the 2- and 3-QL cases, the patterns seem to be composed of two sets of QPI vectors. The inner set is similar to that of thick films (especially for 3 QL), while the shape of the outer one almost resembles the CEC. Besides these unusual QPI vectors caused by point scatterers along the step edge, the strongest QPI features corresponding to the scattering vectors caused by the step-edge reflection show as much brighter spots along the $\bar{\Gamma} - \bar{M}$ direction perpendicular to the step edge, which will be addressed later.

To simulate the QPI patterns for both thick films and those near the critical thickness, we employ the T-matrix approach [28,29] combined with a low energy 4-band effective continuous model in 2D (detailed in section I of [27]). We consider the system including both the top and bottom SSs with the bare Hamiltonian $\mathcal{H}_0 = \int d^2k \langle \psi^\dagger(\mathbf{k}) | H(\mathbf{k}) | \psi(\mathbf{k}) \rangle$, where $|\psi(\mathbf{k})\rangle = (|c_{\mathbf{k}\uparrow 1}\rangle, |c_{\mathbf{k}\downarrow 1}\rangle, |c_{\mathbf{k}\uparrow 2}\rangle, |c_{\mathbf{k}\downarrow 2}\rangle)$. $|c_{\mathbf{k}s\tau}\rangle$ are surface electron states with spin index $s = \uparrow, \downarrow$ and surface index $\tau = 1, 2$ denoting the top and bottom surfaces. On this basis, $H(\mathbf{k}) = \begin{pmatrix} h_1 & t \\ t & h_2 \end{pmatrix}$ with $h_2(\mathbf{k}) = h_1(-\mathbf{k}) + \Delta\mu$. Here $t$, $\Delta\mu$ are the tunneling term and the potential difference between the two surfaces, respectively. Note that the two surfaces ($h_1, h_2$) are inversion symmetric without $\Delta\mu$. For the hexagonal-warped SSs of Bi$_2$Te$_3$, $h_\tau(\mathbf{k}) = \mu_\tau + ak^2 + v(\mathbf{k} \times \boldsymbol{\sigma})_z + \lambda k^3 \cos 3\phi_{\mathbf{k}} \sigma_z$, where $\mu_\tau$, $v$, $\boldsymbol{\sigma}$, and $\phi_{\mathbf{k}}$ are the chemical potential (Dirac energy $E_d$), the fermi velocity, the Pauli matrix, and the azimuthal angle of $\mathbf{k}$, respectively [26]. The square term describes the nonlinearity of the Fermi velocity. The cubic term is the warping term responsible for the strong hexagonally warped SSs in Bi$_2$Te$_3$. Diagonalization of $H(\mathbf{k})$ gives the eigen states $|\tilde{\psi}(\mathbf{k})\rangle$ of $\mathcal{H}_0$, forming four helical SS bands ($|L_+\rangle, |R_+\rangle, |L_-\rangle, |R_-\rangle$) with different spin-momentum helicity (left- or right-handed). Here + and – denote the bands above and below the Dirac point, respectively. The QPI pattern can then be calculated by the Green's function based on the T-matrix as described in sections II and III in [27].

For thick Bi$_2$Te$_3$ films, the top and bottom surfaces are decoupled ($t = 0$). In this case, $(|L_+\rangle, |R_-\rangle)$ and $(|R_+\rangle, |L_-\rangle)$ form the top and bottom SSs as shown in Fig. 2(a), respectively. For the top surface, the two helical bands ($|L_+\rangle, |R_-\rangle$) touch at the Dirac point with their energies satisfy $E_L(\mathbf{k}) = -E_R(\mathbf{k})$ if the square term $ak^2$ is omitted. To calculate the QPI patterns, only the top SSs accessed by the tunneling probe need to be considered. There is only the intra-band ($|L_+\rangle$) scattering in this case (see section IV in [27]). The SS band parameters are adopted from the literature (bulk sample) [5]. The calculated QPI pattern in Fig. 1(d) reproduces nearly perfectly the enhanced scattering vectors $\mathbf{q}$ that are typical for Bi$_2$Te$_3$. The discrepancy in the magnitude of $\mathbf{q}$ might be caused by the mismatch in the surface band parameters (mostly the Fermi velocity) between bulk and thin films. Hence, in the thick-film case, there is only intra-band ($|L_+\rangle$) scattering arcs.

In the presence of tunneling between top and bottom surfaces, a gap opens in the SSs as shown in Fig. 2(b) [16,19,23]. In addition, the 4 helical bands are no longer isolated at each surface. All the bands have contributions from both surfaces. This can be seen in the matrix form of $\hat{U}(\mathbf{k})$ (diagonalizes $H(\mathbf{k})$) in Eqn. S9 of [27]. This enables the inter-band scattering between bands with different helicities where the back scattering is no longer forbidden (Fig. 2(c)). Hence, the calculation of QPI for SSs above the Dirac point now involves inter- and intra-band scattering among $|L_+\rangle$ and $|R_+\rangle$ (see section V of [27]). Note that the intra-band back scattering is still prohibited because the hybridization keeps the helicity of the 4 bands.

We focus on the 1-QL and 3-QL films. Figure 3(a) are the experimental and simulated reciprocal QPI patterns near the step edge of the 1-QL film at 3 different energies. The experimental QPI patterns evolve from a circle into a hexagon with the increasing energy. The calculation catches the essential features of the experimental ones. The parameters used in the calculation are obtained by fitting the band dispersion with the QPI (circles in Fig. 3(b)) and the STS data [27]. Note that the CEC consists of the degenerate $|L_+\rangle$ and $|R_+\rangle$ bands since the $|R_+\rangle$ has top-SS components in the presence of large $t$ in the 1-QL case. The inter-band back-scattering channels are open between $|L_+\rangle$ and $|R_+\rangle$. By comparing the QPI patterns and the corresponding CECs, we see that the QPI patterns come from the back scattering, which makes the patterns mimic the CECs. The QPI is dominated by the back scattering processes. Hence the protection from back scattering in the thick-film case breaks down in the thinnest situation. In the 1-QL case, the absence of intra-band scattering is due to its much-reduced warping effect. The CEC doesn't evolve into a concave shape even at higher energies (see Fig. 3(a) and Fig. S1 [27]), leading to the absence of intra-band scattering arcs.

For 3-QL films, however, things become different compared with the 1-QL case as shown in Fig. 3(c). Inside the hexagon (large $\mathbf{q}$), there appear scattering arcs that become more apparent with the increasing energy (especially at 50 meV), which is consistent with the fact that the warping effect increase with energy. In the simulated patterns, the CECs become more concave and the arcs become sharper with the increasing energy. Figure 3(d) shows the dispersion of SSs (3 QLs) along $\bar{\Gamma} - \bar{M}$.

Nonetheless, although we see in 3-QL films both the hexagon and arcs, by matching vectors in the QPI pattern and those on the CEC (at 50 meV), the weak hexagon is found to come also from intra-band scattering (vectors 3, 4, 5). The back-scattering vector 1 is absent with near-zero intensity. The hexagon actually doesn't match the CEC and is still there in the thick-film case (Fig. S8(a) in [27]). Hence, in the 3-QL case, only the intra-band scattering vectors are identified in the QPI pattern (point scattering). From Fig. 2 we see that even in the presence of a relatively large gap ($\Delta = 0.15$ eV), the top-SS component of $|R_+\rangle$ band drops very rapidly with $k$. For 3-QL films with a nearly diminished SS gap, there is appreciable inter-band scattering only when $k$ approaches zero, which is beyond the scope of point-scattering method. In addition, there are two situations that lead to the discrepancy between experimental and simulated QPI patterns. First, in experimental QPI the scattering pattern at small $q$ is suppressed because the area of pattern is limited in real space. Second, the point-scatterers at the edge suppress scattering vectors near the direction of the edge (see Fig. S7 of [27]).

Hence the QPI based on the point-scattering is not sensitive to the possible presence of tiny inter-surface tunneling. Remember that the step edge induces both point scattering and step-edge reflection. As illustrated in Fig. S7 [27], the step-edge reflection can happen in all positions along the edge and may enhance the reflection vectors on the CEC including both the warping vectors and the possible inter-band back-scattering vectors. The reflection vectors show in the QPI pattern as bright spots along the normal direction of the edge. In the presence of both intra- and inter-band scatterings, there should be two spots along the edge normal with one sitting at the arc position and the other at the tip position of the hexagon. These behaviors are nearly perfectly identified in the reciprocal QPI patterns of 1-4 QLs (Fig. 1(f), 1(h), 1(i) and 1(l)). There are two spots for 2- and 3-QL films. The inner spot sits exactly at the arc position in the 3-QL case. There is only one spot for 1- and 4-QL films. The absence of the inner spot (warping enhanced intra-band scattering) in the 1-QL case is due to the near circular shape of its CEC. Actually, at higher energies the inner spot starts to appear when the CEC become a hexagon (Figs. S1 and S9 of [27]). For 2-QL films, even there is no scattering arcs, the inner spot appears. This is explained by the fact that even the CEC is not concave enough for the appearance of scattering arcs, the intra-band scattering is enhanced around certain vector away from back scattering. As shown in Fig. 4(a), the intra-band reflection probability increases from zero when the vector deviates from the back-scattering case and drops again when the incident and scattered vectors approach the tip positions due to geometric issue. In Fig. 4(b), the QPI is simulated in the presence of both point scattering and step-edge reflection. For 2-QL film at 50 meV, the intra-band scattering appears as an elongated inner spot. In addition, the inter-band back scattering is enhanced by the step-edge reflection and shows as the outer bright spot along the edge normal compared with the diminished intensity of back-scattering vectors along other directions. In contrast to the 1-3 QL cases, in 4-QL films the outer spot is absent (Fig. 1(l) and Fig. S4 of [27]). In addition, the simulation shows that the outer spot appears even in the presence of a tiny tunneling term (the 3-QL case as shown in Fig. S8 of [27]) and disappears in the absence of tunneling (see Fig. 4(c) for the simulated QPI patterns of thick films w/o step-reflection). All the experimental behaviors are summarized in Fig. 4(d). We see that in 4- and 30-QL films (near step edge) the intra-band arcs are absent, probably because the step-edge scatterers are inefficient when the electron-screening becomes strong with thickness. Hence, within the scope of step-reflection approach, there is no inter-surface tunneling in 4-QL films. This sets a lower bound of the critical thickness (4 QLs) for $Bi_2Te_3$ above which the two surfaces are decoupled.

In conclusion, the QPI in ultra-thin $Bi_2Te_3$ films behaves very differently at each thickness due to mainly two factors. The first one is the inter-surface tunneling. The presence of tunneling leads to the appearance of inter-band back-scattering, which get stronger with the decreasing thickness ($t$ increases) and the energy (the SSs mixing decreases with $k$). It appears as the shape of CECs for 1-2 QLs (point scattering) and as bright spot for 1-3 QLs (step reflection). The second one is hexagonal warping, which is found to be very weak in the thinnest case and increases with thickness. The step-edge reflection is more sensitive to the tunneling term by enhancing the possible inter-band back-scattering events. In 4-QL films, the spot corresponding to inter-band back-reflection disappears, indicating the absence of tunneling within the scope of current approach. Furthermore, these complicated scattering behaviors at each thickness are nicely simulated by the T-matrix method based on the 4-band model. Our work establishes the framework for understanding the scattering behaviors in systems with complicated CECs and spin textures.

## ACKNOWLEDGEMENT

The authors acknowledge the supporting from Ministry of Science and Technology of China and National Science Foundation of China (Grants No. 2022YFA1403102, 61804056, 92065102).

**Figure captions**

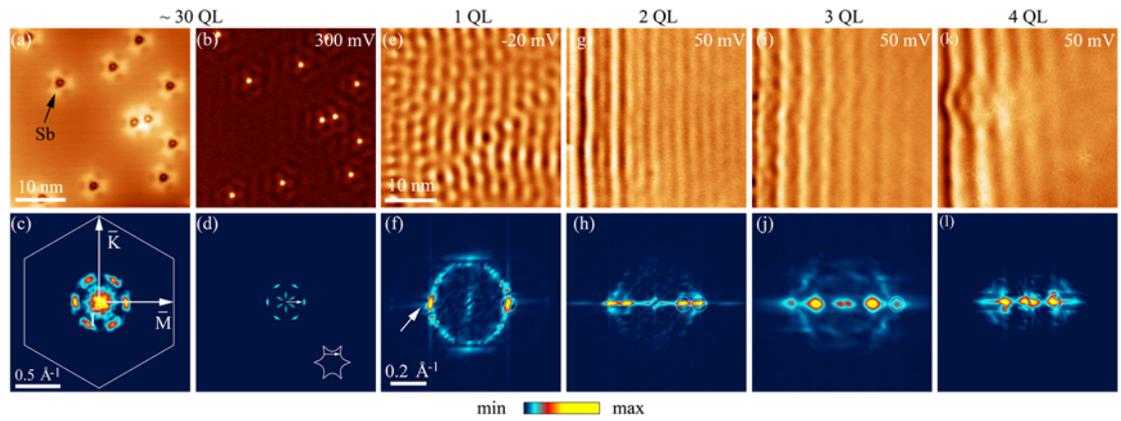

FIG. 1 (color on line). (a) STM image (tunnelling conditions: 300 mV, 50 pA) of a thick Bi$_2$Te$_3$ (111) films with Sb adatoms. (b) Corresponding dI/dV map of (a) at 300 meV. (c) Fast Fourier transform (FFT) image of (b). (d) Calculated quasiparticle interference pattern at 300 meV with parameters [2]: $\lambda = 250$ eV·Å$^3$, $v = 2.55$ eV·Å. (e)-(l) The dI/dV maps near the step edges of 1-4 QL films as well as their FFT images.

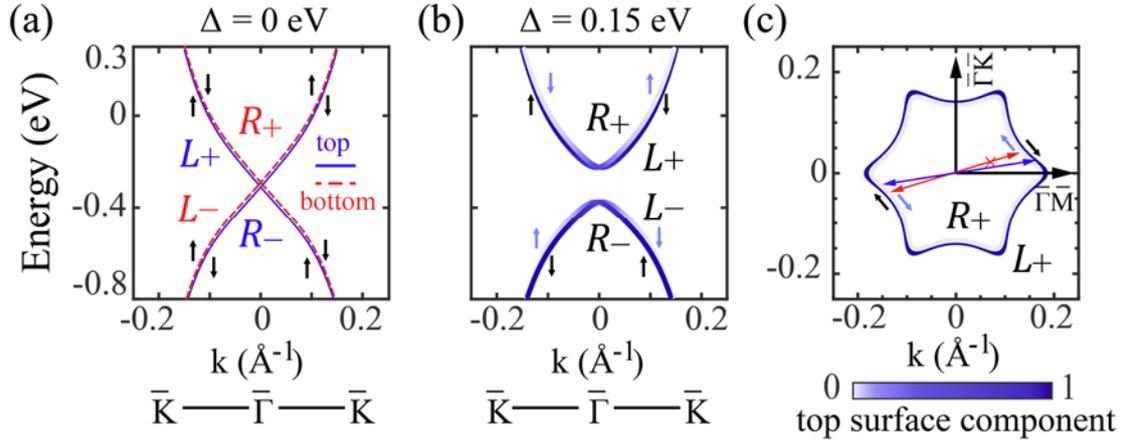

FIG. 2 (color on line) (a) Gapless Dirac conic SSs of top (solid) and bottom (dashed) surfaces calculated by using the parameters of 2 QLs without tunneling ($\Delta = 0$) between two surfaces. A tiny potential difference is introduced to clarify all the four surface bands. The arrows indicated the in-plane spin orientations. (b) The colormap of the gapped SSs with tunneling ($\Delta = 0.15$ eV). The color bar indicates the component coming from the top surface for all the states in these four bands. Here the parabolic term *a* is neglected. (c) The CECs of SSs in (b) at 200 meV.

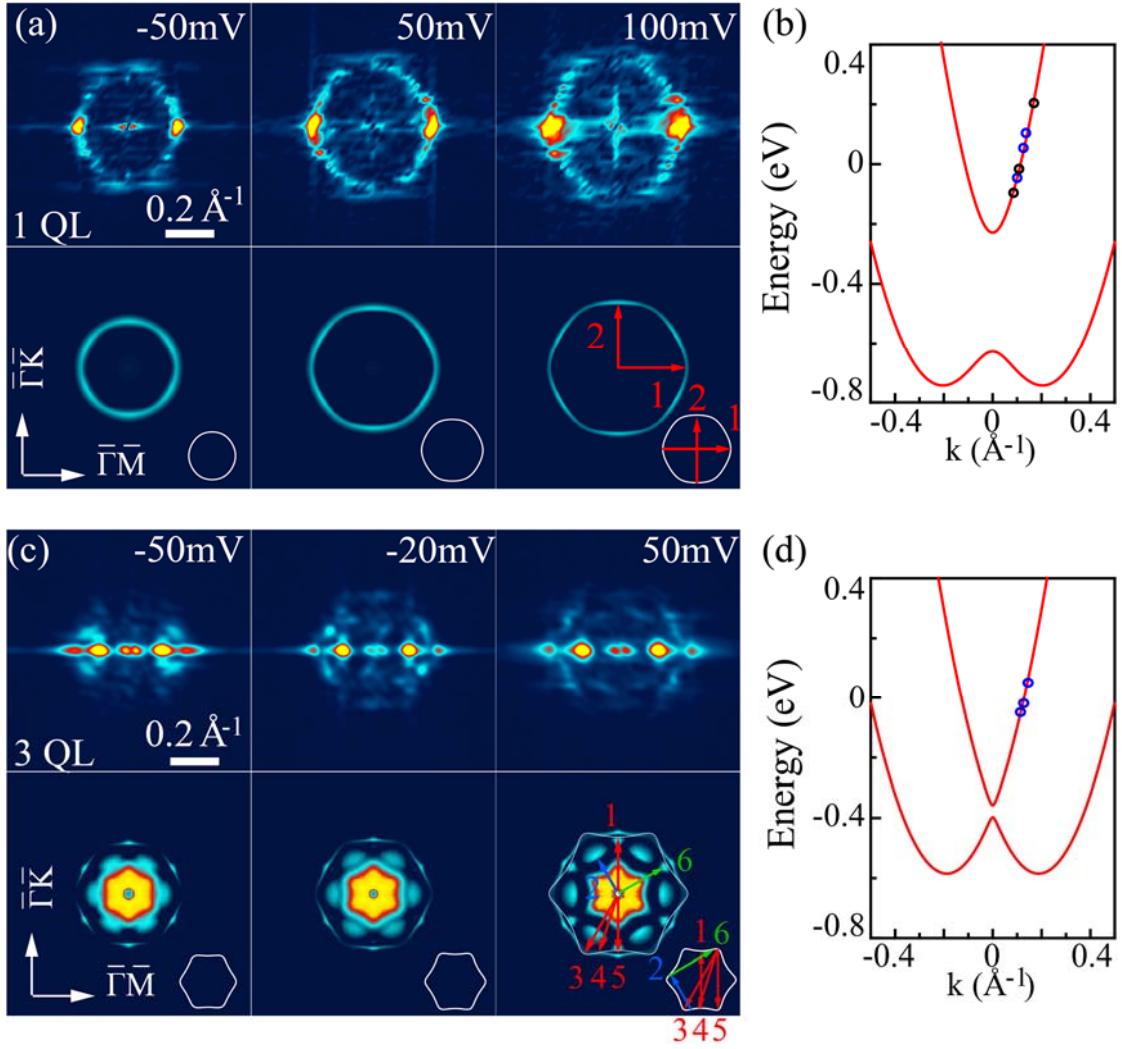

FIG. 3 (color on line) (a) Experimental and simulated QPI patterns of the 1-QL $Bi_2Te_3$ film near the step edge at -50 meV, 50 meV and 100 meV, respectively. (b) The dispersion of SSs along the $\bar{\Gamma} - \bar{M}$ direction, fitted by the experimental data (hollow dots) from -100 meV to 300 meV. The QPI patterns are calculated with parameters: $E_d = -0.42$ eV, $a = 5.83$ eV·Å$^2$, $\lambda = 50$ eV·Å$^3$, $v = 2.55$ eV·Å, $t = 0.2$ eV. Here the gap $\Delta = 2t$. (c) Experimental and simulated QPI patterns ($E_d = -0.38$ eV, $a = 5.83$ eV·Å$^2$, $\lambda = 150$ eV·Å$^3$, $v = 2.2$ eV·Å, $t = 0.02$ eV) of the 3-QL $Bi_2Te_3$ film near the step edge at -50 meV, -20 meV and 50 meV, respectively. (d) The dispersion of SSs along the $\bar{\Gamma} - \bar{M}$ direction, fitted by the experimental data from -50 meV to 200 meV. The hollow dots correspond to the data shown in (c). The gap values for 1-3 QLs are adopted from the literature [19,23].

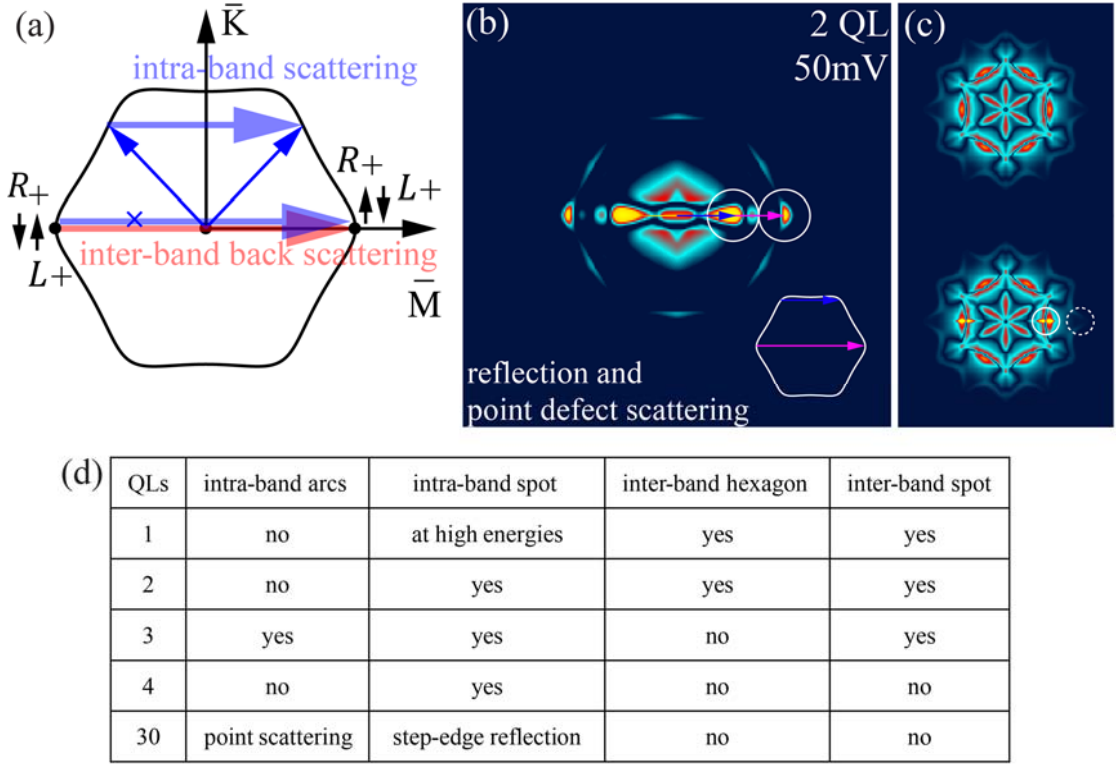

| QLs | intra-band arcs | intra-band spot | inter-band hexagon | inter-band spot |
|---|---|---|---|---|
| 1 | no | at high energies | yes | yes |
| 2 | no | yes | yes | yes |
| 3 | yes | yes | no | yes |
| 4 | no | yes | no | no |
| 30 | point scattering | step-edge reflection | no | no |

FIG. 4 (color on line) (a) The illustration of the inter-band back scattering and the intra-band scattering in the step-edge reflection processes in the reciprocal space for the 2-QL film at 50 meV. (b) The simulated QPI pattern for the 2-QL film at 50 meV in the presence of both step edge reflection and point scatterers sitting at the edge. The parameters used for the calculation are: $E_d = -0.34$ eV, $a = 5.83$ eV·Å², $\lambda = 125$ eV·Å³, $v = 2.27$ eV·Å, $t = 75$ meV. (c) The QPI patterns for thick films without (top) and with step-edge reflection. (d) The summary of QPI features in films with different thicknesses.

.

# References

[1] J. Heil, M. Primke, K. U. Würz, and P. Wyder, Phys. Rev. Lett. **74**, 146 (1995).

[2] A. Weismann, M. Wenderoth, S. Lounis, P. Zahn, N. Quaas, R. G. Ulbrich, P. H. Dederichs, and S. Blügel, Science **323**, 1190 (2009).

[3] M. Chen *et al.*, Sci. Adv. **5**, eaaw3988 (2019).

[4] S. Lounis, P. Zahn, A. Weismann, M. Wenderoth, R. G. Ulbrich, I. Mertig, P. H. Dederichs, and S. Blügel, Phys. Rev. B **83**, 035427 (2011).

[5] Y. L. Chen *et al.*, Science **325**, 178 (2009).

[6] C.-X. Liu, X.-L. Qi, H. Zhang, X. Dai, Z. Fang, and S.-C. Zhang, Phys. Rev. B **82**, 045122 (2010).

[7] Y. Xia *et al.*, Nat. Phys. **5**, 398 (2009).

[8] T. Zhang *et al.*, Phys. Rev. Lett. **103**, 266803 (2009).

[9] P. Roushan *et al.*, Nature **460**, 1106 (2009).

[10] J. Seo, P. Roushan, H. Beidenkopf, Y. S. Hor, R. J. Cava, and A. Yazdani, Nature **466**, 343 (2010).

[11] M. Lang *et al.*, Nano Lett. **13**, 48 (2013).

[12] H.-Z. Lu, J. Shi, and S.-Q. Shen, Phys. Rev. Lett. **107**, 076801 (2011).

[13] M. Liu *et al.*, Phys. Rev. Lett. **108**, 036805 (2012).

[14] H. Wang *et al.*, Sci. Rep. **4**, 5817 (2014).

[15] Y. Zhang *et al.*, Nat. Phys. **6**, 584 (2010).

[16] W.-Y. Shan, H.-Z. Lu, and S.-Q. Shen, New J. Phys. **12**, 043048 (2010).

[17] B. Zhou, H.-Z. Lu, R.-L. Chu, S.-Q. Shen, and Q. Niu, Phys. Rev. Lett. **101**, 246807 (2008).

[18] P. Thalmeier and A. Akbari, Phys. Rev. Res. **2**, 033002 (2020).

[19] T. Förster, P. Krüger, and M. Rohlfing, Phys. Rev. B **93**, 205442 (2016).

[20] Y.-Y. Li *et al.*, Adv. Mater. **22**, 4002 (2010).

[21] Y. Jiang *et al.*, Phys. Rev. Lett. **108**, 016401 (2012).

[22] H.-Z. Lu, W.-Y. Shan, W. Yao, Q. Niu, and S.-Q. Shen, Phys. Rev. B **81**, 115407 (2010).

[23] C.-X. Liu, H. Zhang, B. Yan, X.-L. Qi, T. Frauenheim, X. Dai, Z. Fang, and S.-C. Zhang, Phys. Rev. B **81**, 041307 (2010).

[24] A. A. Zyuzin and A. A. Burkov, Phys. Rev. B **83**, 195413 (2011).

[25] H. Zhang, C.-X. Liu, X.-L. Qi, X. Dai, Z. Fang, and S.-C. Zhang, Nat. Phys. **5**, 438 (2009).

[26] L. Fu, Phys. Rev. Lett. **103**, 266801 (2009).

[27] Additional supplementary text and data are available in the supplementary materials, including the 4-band model, the T-matrix method and the calculation details for both point scattering and step reflection. In addition, the experimental QPI data for 1-4 QLs at other energies are also supplemented.

[28] W.-C. Lee, C. Wu, D. P. Arovas, and S.-C. Zhang, Phys. Rev. B **80**, 245439 (2009).

[29] W.-C. Lee and C. Wu, Phys. Rev. Lett. **103**, 176101 (2009).

# Supplementary information
# Quasiparticle scattering in three-dimensional topological insulators near the thickness limit

Haiming Huang[1†], Mu Chen[2†], Dezhi Song[1], Jun Zhang[1*], Ye-ping Jiang[1*]

*1 Key Laboratory of Polar Materials and Devices (MOE) and Department of Electronics, East China Normal University, Shanghai 200241, China*

*2 Beijing WeLion New Energy Technology Co., Ltd, Beijing 102402, China*

\* Corresponding authors. Email: zhangjun@ee.ecnu.edu.cn, ypjiang@clpm.ecnu.edu.cn

**CONTENTS:**

**I. The 4-band model**

**II. The impurity potential**

**III. The T-matrix method**

**IV. The simulation of QPI in the thick films**

**V. The simulation of QPI in the ultra-thin films**

**VI. The step edge scattering**

**VII. The simulation of QPI in the simultaneous presence of step-edge line reflection and step-edge point scatterers**

**VIII. The effect of step-edge point scatters on the QPI patterns**

**Figs. S1-S9**

**I. The 4-band model**

To simulate the QPI patterns for both thick Bi$_2$Te$_3$ films and those near the critical thickness, we employ the T-matrix approach combined with a low energy 4-band effective continuous model in 2D. We start with the 2×2 effective topological surface Hamiltonian in the spinor basis that describes the hexagonal-warped surface states (SS) of Bi$_2$Te$_3$,

$$h(\mathbf{k}) = \epsilon_D + ak^2 + \hbar v_F(\mathbf{k} \times \boldsymbol{\sigma})_z + \lambda k^3 \cos 3\phi_{\mathbf{k}} \sigma_z, \qquad (1)$$

where $\epsilon_D$, $v_F$, $\boldsymbol{\sigma}$, and $\phi_{\mathbf{k}} = \tan^{-1}(k_y/k_x)$ are the Dirac energy, the fermi velocity, the Pauli matrix, and the azimuthal angle of $\mathbf{k}$, respectively. The spinor basis is $(|c_{\mathbf{k}\uparrow}\rangle, |c_{\mathbf{k}\downarrow}\rangle)$. The square term describes the nonlinearity of the Fermi velocity away from the Dirac point. The cubic term is the additional warping term (with magnitude $\lambda$) responsible for the strong hexagonally warped surface states in Bi$_2$Te$_3$. Substitute the Pauli matrix into Eqn. 1, we get

$$h(k) = \epsilon_D + a'\hat{k}^2 + \epsilon_k \begin{pmatrix} \hat{k}^2 \cos 3\phi_k & -ie^{-i\phi_k} \\ ie^{i\phi_k} & -\hat{k}^2 \cos 3\phi_k \end{pmatrix}. \qquad (2)$$

The $\bar{\Gamma} - \bar{K}$ direction is along $\hat{x}$. $a' = a/(\lambda v_F)^{1/2}$. Here the energy and momentum are in units of $E^* = \hbar v_F k_c$ and $k_c = (v_F/\lambda)^{1/2}$. $\epsilon_k = E^*\hat{k} = \hbar v_F k$ is the isotropic Dirac dispersion and $\hat{k} = k/k_c$.

In the presence of tunneling between top and bottom surface states, we consider the system including both the top and bottom surface states with the bare Hamiltonian

$$\mathcal{H}_0 = \int d^2k \langle \Psi^\dagger(\mathbf{k}) | H(\mathbf{k}) | \Psi(\mathbf{k}) \rangle, \qquad (3)$$

where $|\Psi(\mathbf{k})\rangle = (|c_{\mathbf{k}\uparrow 1}\rangle, |c_{\mathbf{k}\downarrow 1}\rangle, |c_{\mathbf{k}\uparrow 2}\rangle, |c_{\mathbf{k}\downarrow 2}\rangle)$. $|c_{\mathbf{k}s\tau}\rangle$ are surface electron states with spin index $s = \uparrow, \downarrow$ and surface index $\tau = 1, 2$ denoting the top and bottom surfaces. We call it the spin and surface basis. On this basis, $H(\mathbf{k}) = \begin{pmatrix} h_1 & t \\ t & h_2 \end{pmatrix}$ with $h_2(\mathbf{k}) = h_1(-\mathbf{k}) + \Delta\mu$. Here $t$, $\Delta\mu$ is the tunneling matrix and the potential difference between the two surfaces, respectively. Note that the two surfaces are inversion symmetric without $\Delta\mu$. In our work, we choose the condition $\Delta\mu = 0$ because we don't see signatures of splitting between two surfaces.

We first introduce the unitary transformation $S_{\mathbf{k}}$ by which

$$S_k^\dagger h(k) S_k = E(k) \sigma_z. \tag{4}$$

Dispersion of the warped Dirac cone is then expressed as

$$E(k) = \sqrt{(\hbar v_F k)^2 + (\lambda k^3 \cos 3\phi_k)^2} = E^* \hat{k}\sqrt{1 + (\hat{k}^2 \cos 3\phi_k)^2}. \tag{5}$$

The unitary transformation is given by

$$\hat{S}_k = \begin{pmatrix} \cos\frac{\theta_k}{2} & i\sin\frac{\theta_k}{2} e^{-i\phi_k} \\ i\sin\frac{\theta_k}{2} e^{i\phi_k} & \cos\frac{\theta_k}{2} \end{pmatrix}, \tag{6}$$

where $\tan\theta_k = v_F/\lambda k^2 \cos 3\phi_k = 1/\hat{k}^2 \cos 3\phi_k$. The mixed angle $\theta_k$ changes sign six times when the azimuthal angle $\phi_k$ varies from 0 to $2\pi$. For ensuring the continuity of $\theta_k$, we define $\theta_k = \tan^{-1}[1/(\hat{k}^2 \cos 3\phi_k)]$ for $\cos 3\phi_k > 0$ and $\theta_k = \tan^{-1}[1/(\hat{k}^2 \cos 3\phi_k)] + \pi$ for $\cos 3\phi_k < 0$. $\hat{S}_k$ transforms the effective Hamiltonian from the spin basis $(c_{k\uparrow}, c_{k\downarrow})$ to so called "helicity" basis $(|L\rangle_+, |R\rangle_-)$. Here $|L_+\rangle, |R_-\rangle$ correspond to the band above and below the Dirac point and touch at the Dirac point in the massless case in the top surface states. In helicity space, the eigenvectors are the columns of $S_k$, which contain the up and down components of spin. The two eigenvalues $E_L(k)$, $E_R(k)$ satisfy $E_L(k) = -E_R(k)$ if the square term $ak^2$ is omitted.

We now introduce another matrix

$$W_k = \begin{pmatrix} \cos\frac{\psi_k}{2} & 0 & -\sin\frac{\psi_k}{2} & 0 \\ 0 & \sin\frac{\psi_k}{2} & 0 & \cos\frac{\psi_k}{2} \\ \sin\frac{\psi_k}{2} & 0 & \cos\frac{\psi_k}{2} & 0 \\ 0 & \cos\frac{\psi_k}{2} & 0 & -\sin\frac{\psi_k}{2} \end{pmatrix}, \tag{7}$$

with which $H(k)$ can be diagonalized as

$$\tilde{E}(k) = W_k^\dagger \begin{pmatrix} S_k^\dagger & 0 \\ 0 & S_k^\dagger \end{pmatrix} H(k) \begin{pmatrix} S_k & 0 \\ 0 & S_k \end{pmatrix} W_k = U_k^\dagger H(k) U_k$$

$$= \epsilon_D + a'\hat{k}^2 + \begin{pmatrix} \sqrt{E^2(k) + t^2}\kappa_0 & 0 \\ 0 & -\sqrt{E^2(k) + t^2}\kappa_0 \end{pmatrix}. \tag{8}$$

$\kappa$ is Pauli matrices in the helicity basis. Here $\tan\psi_k = \frac{t}{E(k)}$. Then the unitary transformation $U_k$ for $H(k)$ can be expressed as

$$U_k = \begin{pmatrix} S_k & 0 \\ 0 & S_k \end{pmatrix} W_k$$

$$= \begin{pmatrix} \cos\frac{\psi}{2}\cos\frac{\theta}{2} & \sin\frac{\psi}{2}\sin\frac{\theta}{2}ie^{-i\phi} & -\sin\frac{\psi}{2}\cos\frac{\theta}{2} & \cos\frac{\psi}{2}\sin\frac{\theta}{2}ie^{-i\phi} \\ \cos\frac{\psi}{2}\sin\frac{\theta}{2}ie^{i\phi} & \sin\frac{\psi}{2}\cos\frac{\theta}{2} & -\sin\frac{\psi}{2}\sin\frac{\theta}{2}ie^{i\phi} & \cos\frac{\psi}{2}\cos\frac{\theta}{2} \\ \sin\frac{\psi}{2}\cos\frac{\theta}{2} & \cos\frac{\psi}{2}\sin\frac{\theta}{2}ie^{-i\phi} & \cos\frac{\psi}{2}\cos\frac{\theta}{2} & -\sin\frac{\psi}{2}\sin\frac{\theta}{2}ie^{-i\phi} \\ \sin\frac{\psi}{2}\sin\frac{\theta}{2}ie^{i\phi} & \cos\frac{\psi}{2}\cos\frac{\theta}{2} & \cos\frac{\psi}{2}\sin\frac{\theta}{2}ie^{i\phi} & -\sin\frac{\psi}{2}\cos\frac{\theta}{2} \end{pmatrix}. \quad (9)$$

Thus, diagonalization of $H(\boldsymbol{k})$ by a unitary operator $U_k$ gives the eigen states $|\widetilde{\Psi}(\boldsymbol{k})\rangle$ of $\mathcal{H}_0$, forming the four surface state bands $(|L_+\rangle, |R_+\rangle, |L_-\rangle, |R_-\rangle)$ which are the conjugates of columns of the matrix in Eqn. 9. Here $|\widetilde{\Psi}(\boldsymbol{k})\rangle = \widehat{U}_k^\dagger |\Psi(\boldsymbol{k})\rangle$. $(|L_+\rangle, |R_+\rangle, |L_-\rangle, |R_-\rangle)$ denote the helicity basis or the 4-band basis for $H(\boldsymbol{k})$. The system is then transformed from the spin and surface basis to the helicity basis where each band has the same helicity and back scattering is forbidden between states in the same band. Here + and – are the second index denoting the bands above and below the Dirac point, respectively. The simulation of QPI is then based on the four bands $(|L_+\rangle, |R_+\rangle, |L_-\rangle, |R_-\rangle)$.

## II. The impurity potential

In our simulation of QPI in Bi$_2$Te$_3$ films, the impurities are assumed to be short-ranged scatterers that only cause scalar potential scattering. Thus, the scattering matrix $V_{kk'}$ in the impurity-perturbed Hamiltonian

$$\mathcal{H}_{imp} = \int d^2k d^2k' \langle \Psi^\dagger(\boldsymbol{k}) | V_{kk'} | \Psi(\boldsymbol{k}) \rangle \quad (10)$$

can be approximated by a constant $V_{kk'} \approx V_0$. In the 4-band basis of $|\widetilde{\Psi}(\boldsymbol{k})\rangle$, the matrix

$$\widetilde{V}_{kk'} = V_0 \widehat{U}_k^\dagger \widehat{U}_k = \begin{pmatrix} \widetilde{V}_{11} & \widetilde{V}_{12} & \widetilde{V}_{13} & \widetilde{V}_{14} \\ \widetilde{V}_{21} & \widetilde{V}_{22} & \widetilde{V}_{23} & \widetilde{V}_{24} \\ \widetilde{V}_{31} & \widetilde{V}_{32} & \widetilde{V}_{33} & \widetilde{V}_{34} \\ \widetilde{V}_{41} & \widetilde{V}_{42} & \widetilde{V}_{43} & \widetilde{V}_{44} \end{pmatrix}. \quad (10)$$

## III. The T-matrix method

The QPI pattern $\rho(\boldsymbol{q}, \omega)$ in the reciprocal space can then be calculated by the

Green's function based on T-matrix

$$\rho(\boldsymbol{q},\omega) = \int d^2r\, e^{i\boldsymbol{q}\cdot\boldsymbol{r}}\, \rho(\boldsymbol{r},\omega)$$

$$\sim \frac{1}{\pi}\mathrm{Im} \int d^2k\, \mathrm{Tr}[\, U_{\boldsymbol{k}}^\dagger\, G_{\boldsymbol{k},\boldsymbol{k}+\boldsymbol{q}}(\omega)\, U_{\boldsymbol{k}+\boldsymbol{q}}\,], \tag{11}$$

where the Green's Function is in the spin and surface basis $|\psi(\boldsymbol{k})\rangle$,

$$G_{\boldsymbol{k},\boldsymbol{k}'}(\omega) = G_{0\boldsymbol{k}}(\omega)\delta_{\boldsymbol{k},\boldsymbol{k}'} + G_{0\boldsymbol{k}}(\omega)\, T_{\boldsymbol{k},\boldsymbol{k}'}(\omega)\, G_{0\boldsymbol{k}'}(\omega). \tag{12}$$

Here the bare Green's function $[G_{0\boldsymbol{k}}(\omega)]_{ab} = [\omega + i\delta - E_{\boldsymbol{k};a}]^{-1}\delta_{a,b}$, and the T-matrix satisfies

$$T_{\boldsymbol{k},\boldsymbol{k}'}(\omega) = \tilde{V}_{\boldsymbol{k},\boldsymbol{k}'} + \int d^2p\, \tilde{V}_{\boldsymbol{k},\boldsymbol{p}} G_{0\boldsymbol{p}}(\omega) T_{\boldsymbol{p},\boldsymbol{k}'}(\omega). \tag{13}$$

**IV. The simulation of QPI in the thick films**

For thick films, we only need to consider the top surface

$$\mathcal{H}_0 = \int d^2k \langle \psi^\dagger(\boldsymbol{k}) | h(\boldsymbol{k}) | \psi(\boldsymbol{k}) \rangle$$

$$= \int d^2k \langle \tilde{\psi}^\dagger(\boldsymbol{k}) | E(\boldsymbol{k})\sigma_z | \tilde{\psi}(\boldsymbol{k}) \rangle, \tag{14}$$

where $|\psi(\boldsymbol{k})\rangle$, $|\tilde{\psi}(\boldsymbol{k})\rangle$ are in the spinor basis $(|c_{\boldsymbol{k}\uparrow}\rangle, |c_{\boldsymbol{k}\downarrow}\rangle)$ and the helical band basis $(|L_+\rangle, |R_-\rangle)$, respectively. The effective scattering matrix in helicity basis is then

$$\tilde{V}_{\boldsymbol{k},\boldsymbol{k}'} = V_0\, \hat{S}_{\boldsymbol{k}'}^\dagger\, \hat{S}_{\boldsymbol{k}} =$$

$$V_0 \begin{bmatrix} \cos\frac{\theta_k}{2}\cos\frac{\theta_{k'}}{2} + \sin\frac{\theta_k}{2}\sin\frac{\theta_{k'}}{2} e^{i(\phi_k - \phi_{k'})} & i\left(\sin\frac{\theta_k}{2}\cos\frac{\theta_{k'}}{2} e^{-i\phi_k} - \cos\frac{\theta_k}{2}\sin\frac{\theta_{k'}}{2} e^{-i\phi_{k'}}\right) \\ i\left(\sin\frac{\theta_k}{2}\cos\frac{\theta_{k'}}{2} e^{i\phi_k} - \cos\frac{\theta_k}{2}\sin\frac{\theta_{k'}}{2} e^{i\phi_{k'}}\right) & \cos\frac{\theta_k}{2}\cos\frac{\theta_{k'}}{2} + \sin\frac{\theta_k}{2}\sin\frac{\theta_{k'}}{2} e^{i(\phi_{k'} - \phi_k)} \end{bmatrix}. \tag{15}$$

Here the diagonal terms $\tilde{V}_{\boldsymbol{k},\boldsymbol{k}'}^{11}$ and $\tilde{V}_{\boldsymbol{k},\boldsymbol{k}'}^{22}$ in Eq(15) are intra-band scattering of the upper ($|L_+\rangle$) or lower ($|R_-\rangle$) eigen band respectively, nondiagonal terms $\tilde{V}_{\boldsymbol{k},\boldsymbol{k}'}^{12}$ and $\tilde{V}_{\boldsymbol{k},\boldsymbol{k}'}^{21}$ are the inter-band scattering between upper to lower helical bands. The inter-band scattering can be neglected in the elastic case here (note the $\delta_{a,b}$ function in the Green's function). $\tilde{V}_{\boldsymbol{k},\boldsymbol{k}'}^{11}, \tilde{V}_{\boldsymbol{k},\boldsymbol{k}'}^{22}$ are modulated by factor $(\phi_k - \phi_{k'})$ in momentum space, especially when $|\phi_k - \phi_{k'}| \to \pi$ where these amplitudes are strongly suppressed. This forbids the intra-band backscattering.

## V. The simulation of QPI in the ultra-thin films

For ultra-thin films, we need to consider both surfaces

$$\mathcal{H}_0 = \int d^2k \langle \Psi^\dagger(\mathbf{k})|H(\mathbf{k})|\Psi(\mathbf{k})\rangle$$
$$= \int d^2k \langle \widetilde{\Psi}^\dagger(\mathbf{k})|E(\mathbf{k})\kappa_0 \otimes \tau_z|\widetilde{\Psi}(\mathbf{k})\rangle, \quad (16)$$

where $|\Psi(\mathbf{k})\rangle$, $|\widetilde{\Psi}(\mathbf{k})\rangle$ are in the spinor basis $(|c_{k\uparrow 1}\rangle, |c_{k\downarrow 1}\rangle, |c_{k\uparrow 2}\rangle, |c_{k\downarrow 2}\rangle)$ and the helical band basis $(|L_+\rangle, |R_+\rangle, |L_-\rangle, |R_-\rangle)$, respectively. $\tau = \pm$ denotes the band index above and below the Dirac point. The effective scattering matrix in helicity basis is then the Eqn. 10.

The $2 \times 2$ block in the main diagonal of the $4 \times 4$ $\widetilde{V}_{k,k'}$

$$\begin{pmatrix} \widetilde{V}_{11} & \widetilde{V}_{12} \\ \widetilde{V}_{21} & \widetilde{V}_{22} \end{pmatrix} \quad (17)$$

describes the intra- and inter-band scattering among the bands above the Dirac point $|L_+\rangle$ and $|R_+\rangle$. We have

$$\widetilde{V}_{11} = \cos\left(\frac{\psi_k}{2} - \frac{\psi_{k'}}{2}\right)\left[\cos\frac{\theta_k}{2}\cos\frac{\theta_{k'}}{2} + \sin\frac{\theta_k}{2}\sin\frac{\theta_{k'}}{2}e^{-i(\phi_k-\phi_{k'})}\right];$$

$$\widetilde{V}_{12} = \sin\left(\frac{\psi_k}{2} + \frac{\psi_{k'}}{2}\right)\left[\cos\frac{\theta_k}{2}\sin\frac{\theta_{k'}}{2}ie^{-i\phi_{k'}} - \sin\frac{\theta_k}{2}\cos\frac{\theta_{k'}}{2}ie^{-i\phi_k}\right];$$

$$\widetilde{V}_{21} = \sin\left(\frac{\psi_k}{2} + \frac{\psi_{k'}}{2}\right)\left[\cos\frac{\theta_k}{2}\sin\frac{\theta_{k'}}{2}ie^{i\phi_{k'}} - \sin\frac{\theta_k}{2}\cos\frac{\theta_{k'}}{2}ie^{i\phi_k}\right];$$

$$\widetilde{V}_{22} = \cos\left(\frac{\psi_k}{2} - \frac{\psi_{k'}}{2}\right)\left[\cos\frac{\theta_k}{2}\cos\frac{\theta_{k'}}{2} + \sin\frac{\theta_k}{2}\sin\frac{\theta_{k'}}{2}e^{i(\phi_k-\phi_{k'})}\right]. \quad (18)$$

Regarding the issue of back scattering, we take $\phi_{k'} = \phi_k + \pi$ ($\theta_k + \theta_{k'} = \pi$). We then have

$$\widetilde{V}_{11} = \cos\left(\frac{\psi_k}{2} - \frac{\psi_{k'}}{2}\right)\left[\cos\frac{\theta_k+\theta_{k'}}{2}\right] = 0;$$

$$\widetilde{V}_{12} = -i\sin\left(\frac{\psi_k}{2} + \frac{\psi_{k'}}{2}\right)\left[\sin\frac{\theta_k+\theta_{k'}}{2}\right]e^{-i\phi_k} = -i\sin\left(\frac{\psi_k}{2} + \frac{\psi_{k'}}{2}\right)e^{-i\phi_k};$$

$$\widetilde{V}_{21} = -i\sin\left(\frac{\psi_k}{2} + \frac{\psi_{k'}}{2}\right)\left[\sin\frac{\theta_k+\theta_{k'}}{2}\right]e^{i\phi_k} = -i\sin\left(\frac{\psi_k}{2} + \frac{\psi_{k'}}{2}\right)e^{i\phi_k};$$

$$\widetilde{V}_{22} = \cos\left(\frac{\psi_k}{2} - \frac{\psi_{k'}}{2}\right)\left[\cos\frac{\theta_k+\theta_{k'}}{2}\right] = 0.$$

Because $\tan\psi_k = \frac{t}{E(\mathbf{k})}$, $\widetilde{V}_{12}$, $\widetilde{V}_{21}$ are none-zero only when $t \neq 0$. Hence, the inter-band back scattering happens only when there is tunneling between the top and bottom surfaces. In addition, the intra-band $\tilde{V}_{11}$, $\tilde{V}_{22}$ are always zero. It's because the tunneling between surfaces preserves the time reversal symmetry.

## VI. The step edge reflection

For thick films, the step-edge reflection only leads to one scattering vector corresponding to the warping enhanced scattering along $\bar{\Gamma} - \bar{M}$ (see Fig. S6). For ultra-thin films where the top and bottom surfaces hybridize, there appears inter-band back scattering. This is because the $|R_+\rangle$ band (originally belongs to the bottom surface) acquires the top surface state components in the presence of inter-surface tunneling as illustrated in Fig. 2. The inter-band back reflection leads to the appearance of the second scattering vector at larger $q$ in the reciprocal QPI patterns for 2- and 3-QL films (Fig. 1).

In the 1-QL case, the intra-band reflection vector is absent because of its near circular shape of CEC. Because the intra-band back reflection is forbidden, the possible intra-band reflection only happens away from the back reflection situation (see Fig. 4(a) for the 2-QL case). There is no geometric enhancement of the scattering vectors in the case of a near-circular CEC. Hence, for 1-QL films, there is only one bright spot corresponding to the inter-band back scattering as shown in Fig. 1(f).

In the 2-QL case, although the CEC is not concave enough to ensure the presence of parallel CEC sectors (to favor nesting vectors), the hexagonal shape already leads to the enhancement of certain scattering vector that could lead to the appearance of spot-like feature in the QPI pattern (see the simulated QPI for 2-QL films in the step-edge reflection case in Fig. 4(b)). This is because, in the hexagonal CEC case, the intra-band reflection factor increases gradually when the incident and reflected wave vectors deviate from the $\bar{\Gamma} - \bar{M}$ direction (the intra-band back reflection along this direction is forbidden), while the geometric factor almost keeps the same contrary to the circular CEC case where the geometric factor decreases rapidly. Hence, for 2-QL films, there are two bright spots in the $\bar{\Gamma} - \bar{M}$ direction corresponding to the inter-band back scattering (outer one) and the intra-band reflection (Fig. 1(h)), respectively.

For 3-QL films, the outer bright spot is very weak. It's reasonable because of the strong hexagonal warping in thicker films, where the geometric factor decreases rapidly at the tips along $\bar{\Gamma} - \bar{M}$ directions in the snow-flake like CEC.

## VII. The simulation of QPI in the simultaneous presence of step-edge line reflection and step-edge point scatterers

The calculation of the QPI pattern for step-edge reflection case can be conducted as follows. We modify the matrix elements in Eqn. 11 by the coefficient (a δ function) $\delta((k_x + k'_x)^2 + (k_y - k'_y)^2)$ with a momentum broadening equal to that coming from the energy broadening. This provides the restriction that the scattering belongs to the reflection process (Fig. S7). The simulated QPI patterns for the 2-QL and 3-QL cases are shown in Fig. 4(b) and Fig. S8(d), respectively, which captures the main features in Figs. 1(h) and 1(j).

## VIII. The effect of step-edge point scatters on the QPI patterns

In the main text, the step-edge scatters are treated as ideal point scatterers that can scatter the incident wave in any directions. In fact, the incident wave can only approach the step-edge point scatter in the $k_x < 0$ condition ($Bi_2Te_3$ is in the x > 0 region) and the scattered wave can only be in the $k'_x > 0$ condition (Figs. S7(c) and S7(d)). Hence, the scattering vector *q* near the $\bar{\Gamma} - \bar{K}$ direction (parallel to $\hat{y}$) should be diminished because it requires the condition that $k_x \approx k'_x$ which is not satisfied in most scattering events. By imposing the above restrictions, the QPI patterns in the manuscript that show diminished features along the direction of the step edge can be simulated (not shown).

**Figure captions**

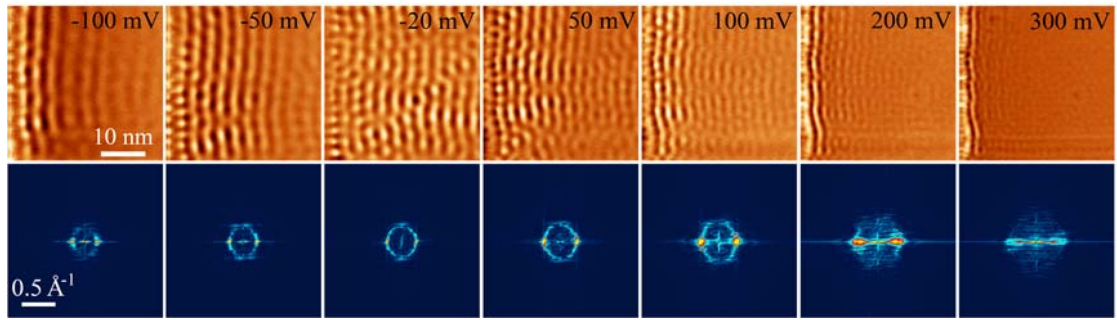

FIG. S1 (color on line). dI/dV maps (35 × 35 nm$^2$) and the corresponding FFT images of 1-QL Bi$_2$Te$_3$ near the step edge at different sample voltages.

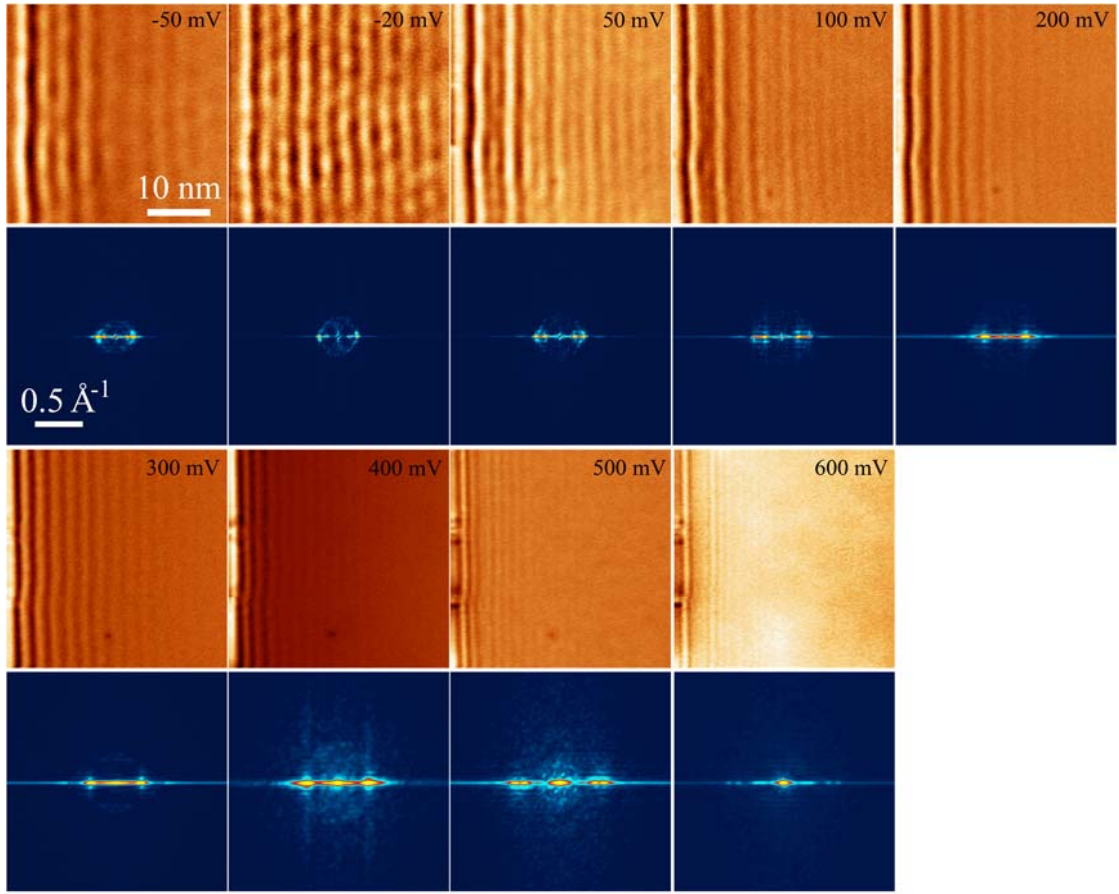

FIG. S2 (color on line). dI/dV maps (35 × 35 nm$^2$) and the corresponding FFT images of 2-QL Bi$_2$Te$_3$ near the step edge at different sample voltages.

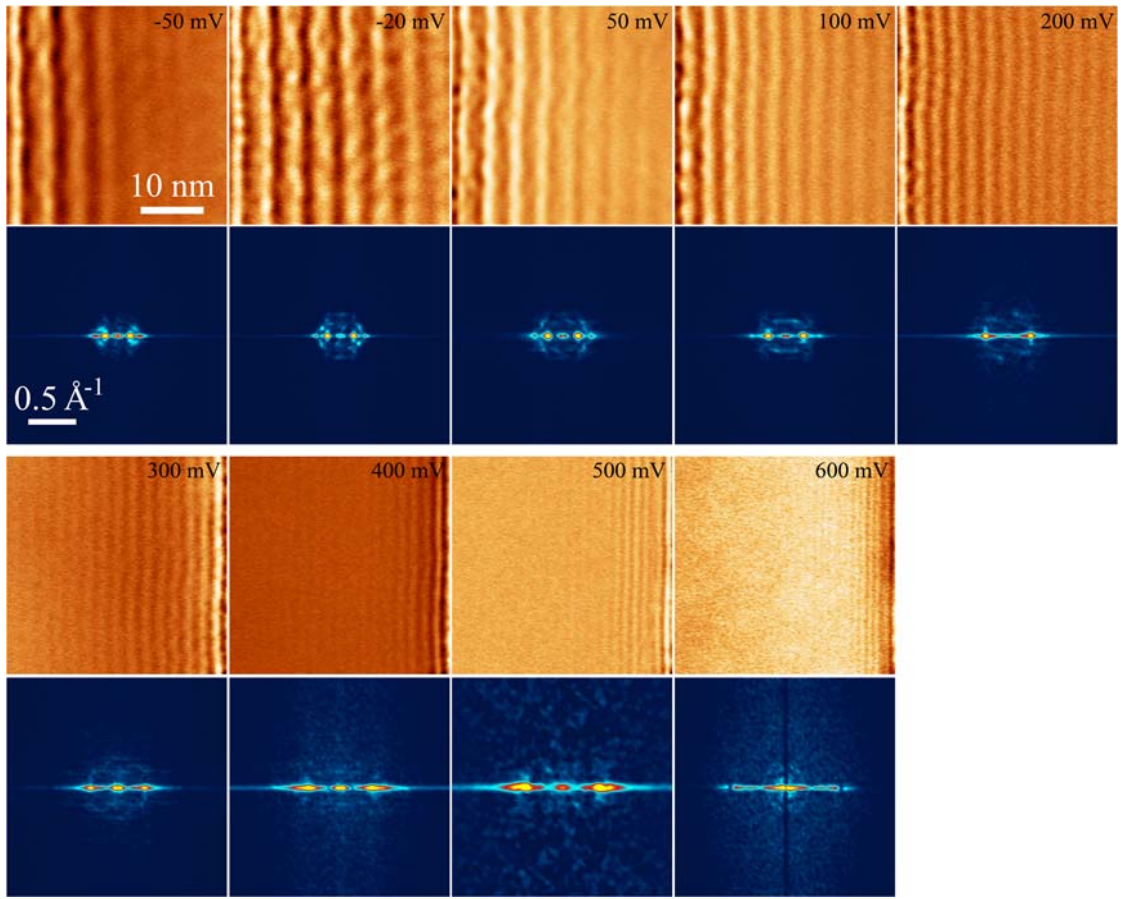

FIG. S3 (color on line). dI/dV maps (35 × 35 nm$^2$) and the corresponding FFT images of 3-QL Bi$_2$Te$_3$ near the step edge at different sample voltages.

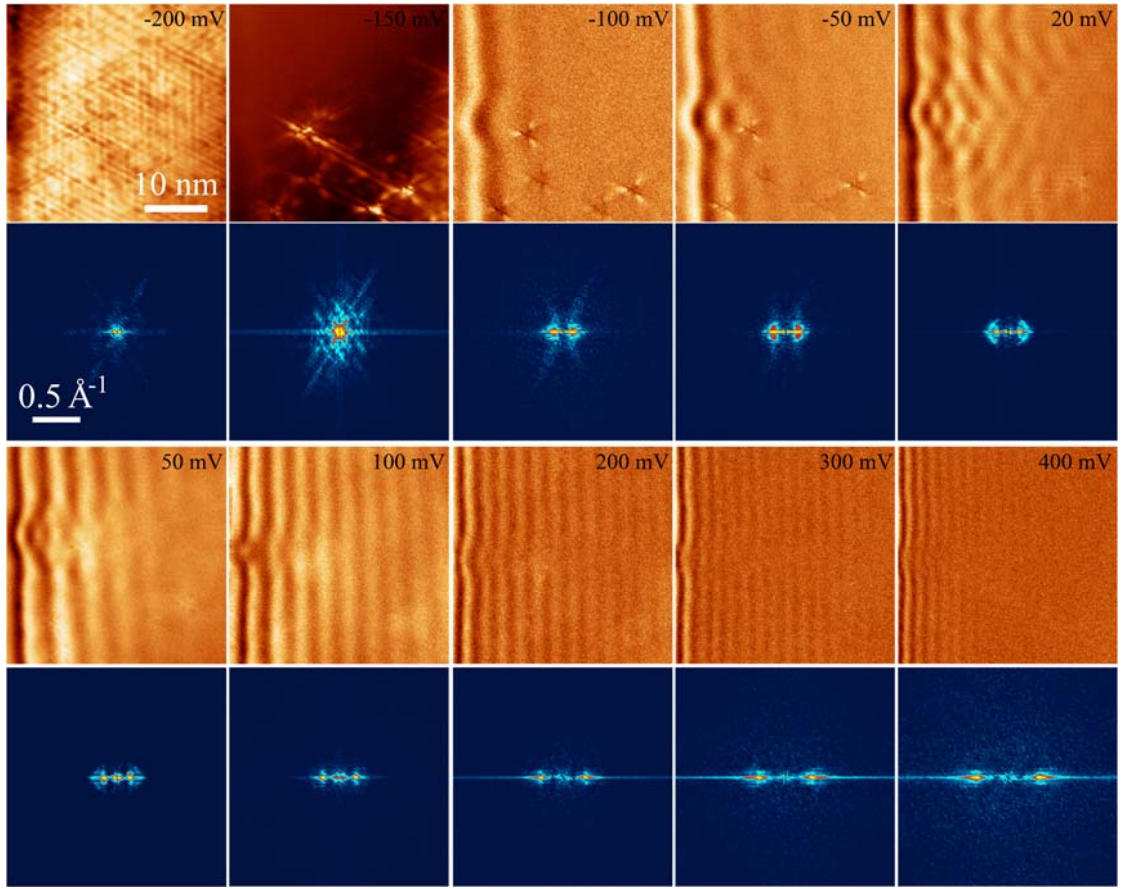

FIG. S4 (color on line). dI/dV maps (35 × 35 nm$^2$) and the corresponding FFT images of 4-QL Bi$_2$Te$_3$ near the step edge at different sample voltages.

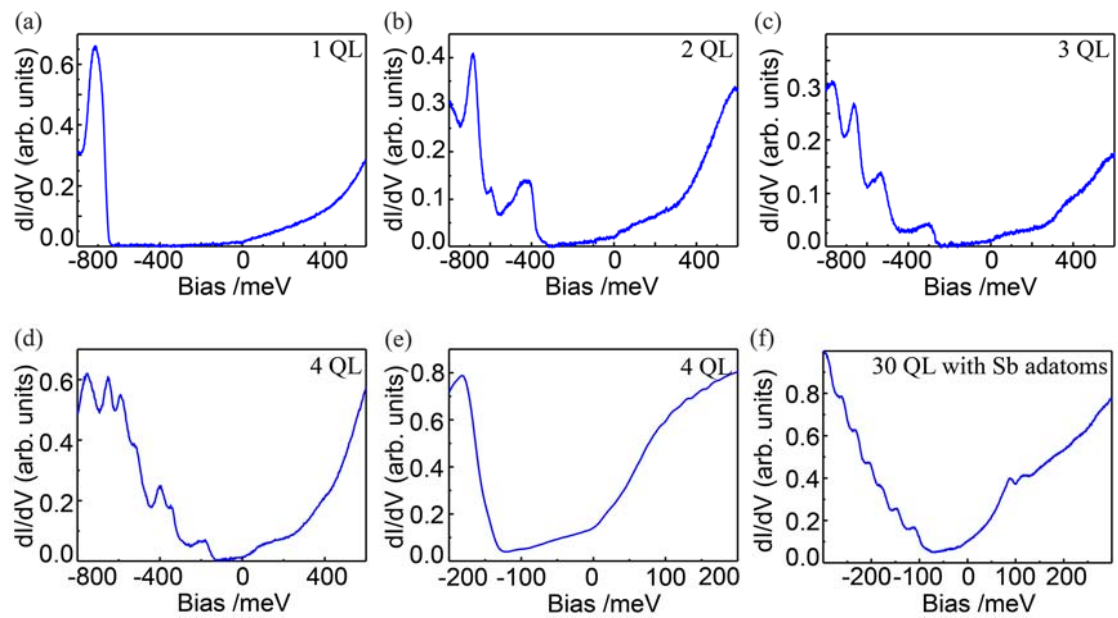

FIG. S5 (color on line) (a)-(f) The dI/dV spectra taken at 1-4 QL $Bi_2Te_3$ films and the 30-QL film with Sb adatoms.

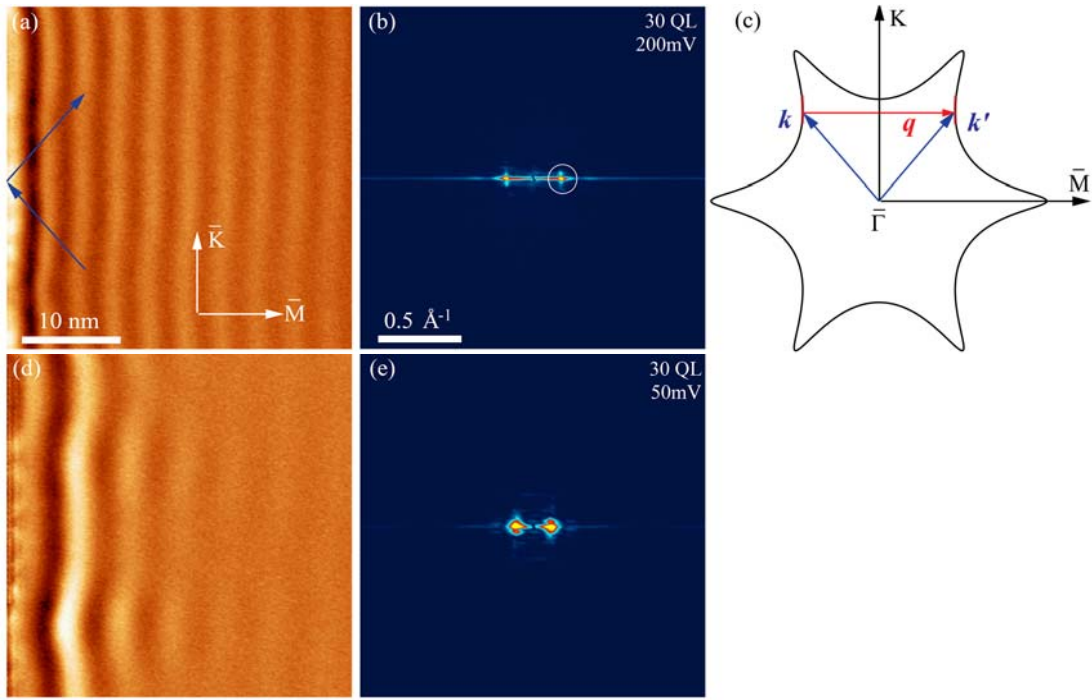

FIG. S6 (color on line) (a) and (b) The dI/dV map and the corresponding FFT image of the 30-QL $Bi_2Te_3$ film near the step edge at 200 meV. (c) The reflection process illustrated in the reciprocal space. (d) and (e) The dI/dV map and the corresponding FFT image of the 30-QL $Bi_2Te_3$ film near the step edge at 50 meV.

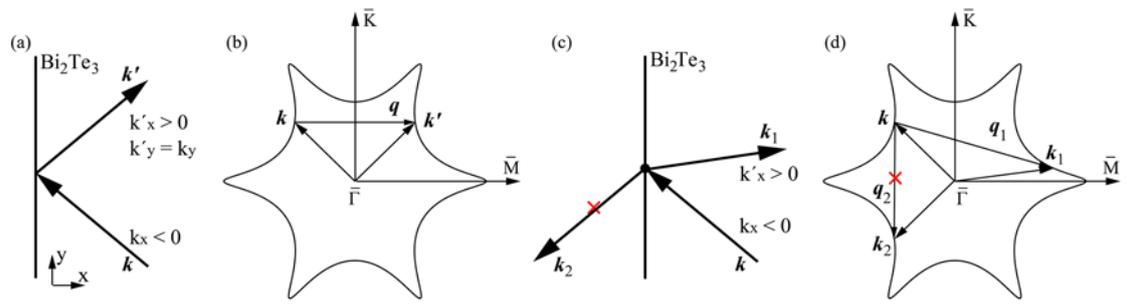

FIG. S7 (color on line) (a) and (b) The schematic step edge reflection in the real and reciprocal space. (c) and (d) The schematic illustration of the point-defect scattering near the step edge in the real and reciprocal space. The red cross indicates the forbidden scattering vector in this situation.

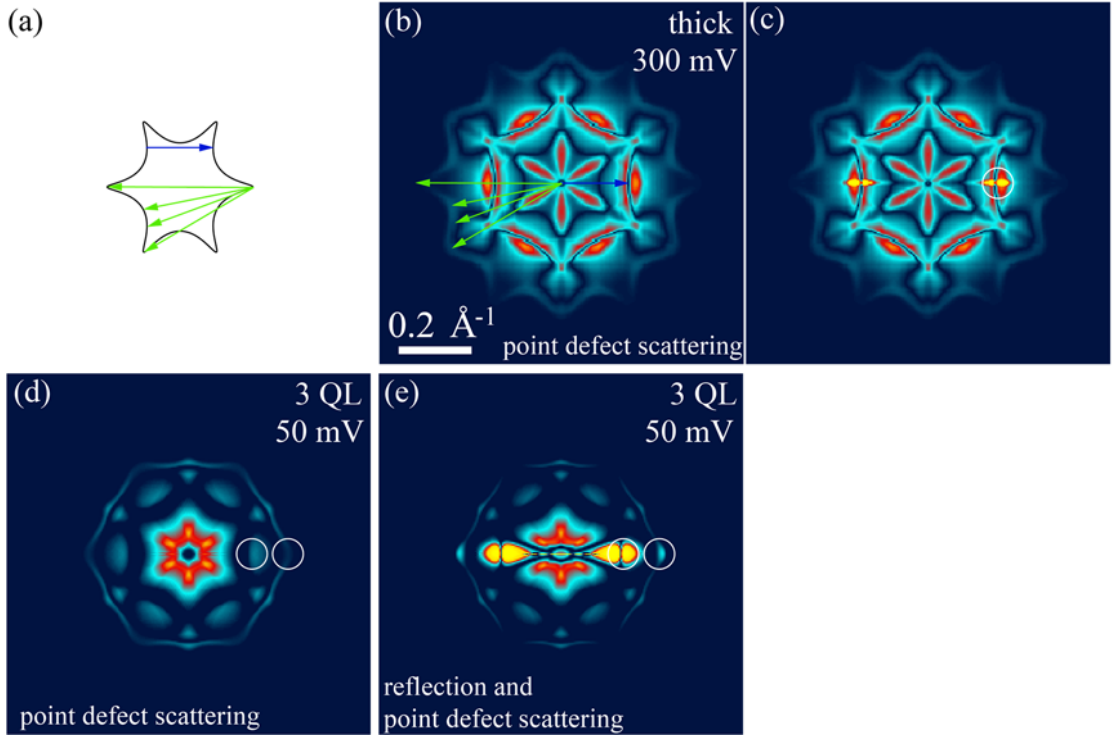

FIG. S8 (color on line) (a) The CEC for thick films at 300 meV. (b) The simulated QPI pattern (point defect scattering) for thick films at 300 meV. The color scale is modified to show patterns that are very weak compared with the pattern of warping-induced scattering. The vectors denoting the strong warping-induced scattering and those weak intra-band patterns are sketched. (c) The simulated QPI pattern for thick films with both point scattering and step-edge reflection. (d) (e) The simulated QPI patterns with only point defect scattering and with the addition of step edge reflection for the 3-QL film at 50 meV, respectively. The parameters used for the calculation are: $E_d = -0.38$ eV, $a = 5.83$ eV·Å$^2$, $\lambda = 150$ eV·Å$^3$, $v = 2.55$ eV·Å, $t = 0.02$ eV.

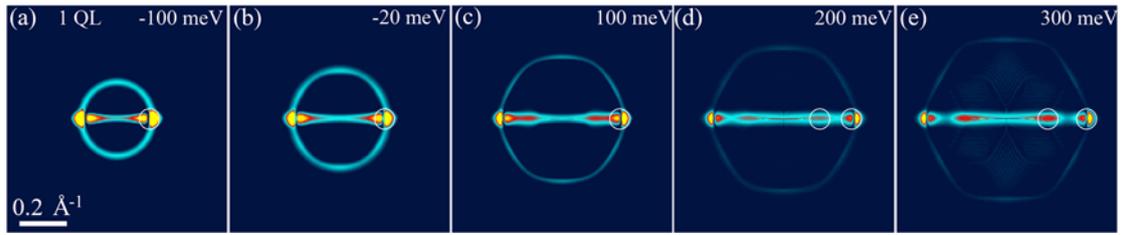

FIG. S9 (color on line) The simulated QPI patterns with both point scattering and step-edge reflection for the 1-QL film at -100, -20, 100, 200, 300 meV, respectively.